\documentclass[reprint,nofootinbib,amsmath,amssymb,fontenc aps]{revtex4-1}
\usepackage{graphicx,dcolumn,bm,hyperref,float}
\usepackage{anyfontsize}
\usepackage{color}
\usepackage{xcolor}

\newcommand{\beq}{\begin{equation}}
\newcommand{\eeq}{\end{equation}}
\newcommand{\bea}{\begin{eqnarray}}
\newcommand{\eea}{\end{eqnarray}}
\newcommand{\bef}{\begin{figure}}
\newcommand{\eef}{\end{figure}}

\newcommand{\mpl}{M_{\mbox{\tiny{Pl}}}}

\newcommand{\Vt}{V_{\mbox{\tiny{tidal}}}}
\newcommand{\pf}{\gamma}
\newcommand{\phidg}{\phi_{\mbox{\tiny{DG}}}}
\newcommand{\phiH}{\phi_{\mbox{\tiny{H}}}}
\newcommand{\bphidg}{\bar{\phi}_{\mbox{\tiny{DG}}}}
\newcommand{\rhonfw}{\rho_{\mbox{\tiny{H}}}}
\newcommand{\rhoH}{\rho_{\mbox{\tiny{H,ave}}}}
\newcommand{\omegadim}{\omega_{\mbox{\tiny{dim}}}}
\newcommand{\bc}{B}

\begin{document}

\title{Quantum Tunneling of Ultralight Dark Matter Out of Satellite Galaxies}

\author{Mark P.~Hertzberg$^{1,2,3,4}$}
\email{mark.hertzberg@tufts.edu}
\author{Abraham Loeb$^2$}
\email{aloeb@cfa.harvard.edu}
\affiliation{$^1$Institute of Cosmology, Department of Physics and Astronomy, Tufts University, Medford, MA 02155, USA
\looseness=-1}
\affiliation{$^2$Department of Astronomy, Harvard University, 60 Garden Street, Cambridge, MA 02138, USA
\looseness=-1}
\affiliation{$^3$Department of Physics, Harvard University, Cambridge, MA 02138, USA
\looseness=-1}
\affiliation{$^4$Department of Physics, Brown University, Providence, RI, 02912, USA
\looseness=-1}

\begin{abstract}
The idea of ultralight scalar (axion) dark matter is theoretically appealing and may resolve some small-scale problems of cold dark matter; so it deserves careful attention. In this work we carefully analyze tunneling of the scalar field in dwarf satellites due to the tidal gravitational force from the host halo. The tidal force is far from spherically symmetric; causing tunneling along the axis from the halo center to the dwarf, while confining in the orthogonal plane. We decompose the wave function into a spherical term plus higher harmonics, integrate out angles, and then numerically solve a residual radial Schr\"odinger-Poisson system. By demanding that the core of the Fornax dwarf halo can survive for at least the age of the universe places a bound on the dark matter particle mass $2\times 10^{-22}\,\mbox{eV}\lesssim m\lesssim 6\times 10^{-22}\,$eV. Interestingly, we show that if another very low density halo is seen, then it rules out the ultralight scalar as core proposal completely. Furthermore, the non-condensed particles likely impose an even sharper lower bound. We also determine how the residual satellites could be distributed as a function of radius.
\end{abstract}

\maketitle


{\em Introduction}.---
The mystery of the nature of dark matter (DM) remains a central puzzle in modern cosmology. Some popular candidates for the DM, such as weakly-interacting massive particles, have faced increased pressure over the years through the lack of direct detection. Perhaps the most well-motivated remaining possibility is a kind of ``axion" \cite{Peccei:1977hh,Weinberg:1977ma,Wilczek:1977pj,Preskill:1982cy,Abbott:1982af,Dine:1982ah}. In its original form, it is a new light scalar ($\phi$) that is postulated to carry a shift symmetry. It therefore has no couplings to the Standard Model (SM) at dimension 4, but can couple to gluons through the dimension 5 operator $\sim\phi\, G\, \tilde{G}$. Careful analysis reveals it resolves the strong CP problem and also picks up a small but nonzero mass and is a DM candidate. Beyond this, there are many related possibilities; including string inspired axions, many of which are typically ``ultralight", perhaps many orders of magnitude lighter than a neutrino. Some axions acquire a mass through gravitational instantons with action $S$ on the order $m\sim \mpl\Lambda\,\exp(-S/2)/F$, where $\mpl=1/\sqrt{8\pi G}$ is the Planck mass and $F$ is some high symmetry breaking scale. Typical instanton actions have $S\sim 2\pi/\alpha$, so for $\alpha\sim 1/25$ as anticipated in unification, the axion's mass would be incredibly small.

Independently, observations have accumulated suggesting that the vanilla cold dark matter (CDM) paradigm may need refining. These observations include the missing satellites problem, suppression of small scalar power, and the presence of cores instead of cusps at the centers of galaxies \cite{Bullock:2017xww}. All of this could conceivably point towards the need for an ultralight scalar as the DM, with a mass perhaps on the order of $m\sim 10^{-22}$\,eV or so. For such a mass, the corresponding de Broglie wavelength for virialized scalars in the galaxy, with velocities $v\sim 10^{-3}c$, would be $\lambda_{dB}\sim$\,kpc; on the order the size of galactic cores. For an ultralight axion formed from the misalignment mechanism (field initially displayed away from the vacuum and then rolls when $H<m$), its relic abundance can be readily shown to be $\Omega_\phi\sim 0.2\,(F/(10^{17}\,\mbox{GeV})^2\,(m/(10^{-22}\,\mbox{eV})^{1/2}$. So the required abundance $\Omega\sim 0.25$ is achieved for $m\sim 10^{-22}$\,eV and $F\sim 10^{17}$\,GeV, which is compatible with the above relation between $m$ and $F$; this is the ``fuzzy miracle" \cite{Bachlechner:2019vcb}. 

In this work we carefully analyze a novel phenomenon that puts serious pressure on the viability of ultralight scalars.  Since these scalars are so light and have such a huge de Broglie wavelength, one may wonder if there can be interesting quantum behavior on the relevant macroscopic scales. One such interesting phenomenon is the formation of solitons at the cores of galaxies, which is observationally enticing. On the other hand, another possible phenomenon is quantum tunneling, which may cause satellite galaxies and dwarfs to deplete their abundance over time and essentially disappear. If this happens too quickly, it would prevent ultralight axions from providing any dwarfs from existing today. The tunneling phenomena can occur because there are two gravitational effects at play. On the one hand, a dwarf galaxy will tend to hold the DM in with its self gravity. On the other hand, the host halo provides a tidal potential that leads to a local maximum in the effective potential. While classical particles would always remain within the corresponding tidal radius, quantum particles with very large de Broglie wavelengths could tunnel across this barrier. 

In previous work \cite{Loeb:2022otx}, one of us studied this tunneling, finding a bound on the scalar's mass and it was also studied in the Appendix of Ref.~\cite{Hui:2016ltb}. In these previous works a simplifying assumption of spherical symmetry was used to describe the effective potential. Detailed numerical analysis of the problem was performed in Ref.~\cite{Du:2018qor}.
In this work, we will develop the analytics and compare to data. We will describe the high asphericity of the tidal potential; it is destabilizing along the axis from halo center to dwarf, but stabilizing in the orthogonal plane. We will also generalize the analysis to a proper treatment of the halo's potential, which deviates from a $1/r$ potential within the halo. We use this analysis to place both a lower and upper bound on the axion mass, as well as a bound on dwarf's core densities to avoid tunneling.

The outline of our paper is as follows. 
We first recap the standard way to study light bosons at very high occupancy, interacting gravitationally, by deriving the Schr\"odinger-Poisson system. For an orbiting dwarf galaxy, we move to a rotating coordinate system. We then break up the gravitational term into a self-gravitational piece for a dwarf galaxy and an external piece from the host halo which is decomposed into a tidal term to second order. 
We then break up the wave function into a spherical piece plus spherical harmonics, then integrate out angles. By determining the inner and outer asymptotic behavior, a residual radial differential equation is solved numerically for the profile and the tunneling rate. 
We also compute how the tunneling is altered when the satellite is within the halo, finding a maximal tunneling rate near the halo's radius.

{\em Schr\"odinger Poisson System}.---
We begin by considering a massive scalar $\phi$, minimally coupled to gravity. The full relativistic action is given by (signature + - - - and units $\hbar=c=1$)
\beq
S=\int \!d^4x\,\sqrt{-g}\left[{\mathcal{R}\over16\pi G}+{1\over 2}g^{\mu\nu}\partial_\mu\phi\partial_\nu\phi-{1\over2}m^2\phi^2\right].
\eeq
(A more precise treatment for the axion involves a periodic potential, such as $V(\phi)=m^2F^2(1-\cos(\phi/F))$; but expanded to quadratic order, gives the above $V(\phi)={1\over2}m^2\phi^2$ which suffices in the galaxy where densities are low.) 
A typical axion has incredibly small couplings to matter and so they can be ignored when considering motion in the galaxy, as we shall do here.  

Within a galaxy, the matter has approximately virialized to speeds $v\sim 10^{-5}-10^{-3}$ and so a non-relativistic approximation may be employed. To capture this, we make the standard decomposition of the rapidly oscillating $\phi$ in terms of a slowly varying ``wave function" $\psi$ as
\beq
\phi(t,{\bf x})={1\over\sqrt{2m}}\left[e^{-imt}\psi(t,{\bf x})+e^{imt}\psi^*(t,{\bf x})\right].
\eeq 
In the high occupancy regime, this $\psi$ is more precisely the non-relativistic (Schr\"odinger) {\em field}, rather than the many particle wave function. But since it captures the coherent quantum physics of the underlying particles, the name ``wave function" is suggestive. We note that from the particle point of view, the upcoming tunneling is indeed a {\em quantum} phenomenon which can be called ``quantum tunneling", while from the field point of view it is captured by {\em classical} field theory. 

By inserting this decomposition of $\phi$ into the above action, operating in Newtonian gauge, integrating out the rapidly varying parts, we can obtain a non-relativistic (more precisely, a Galilean relativistic) effective action
\beq
S = \int d^4x\left[i\,\dot\psi\,\psi^*-{\nabla\psi^*\!\cdot\!\nabla\psi\over 2m}-m\phi_N\psi^*\psi-{(\nabla\phi_N)^2\over 8\pi G}\right],
\eeq
(one takes the real part). 
The corresponding equations of motion are the Sch\"odinger-Poisson system
\bea
&&i\,\dot\psi=-{\nabla^2\psi\over 2m}+m\,\phi_N\,\psi,\,\,\,\,\,\,
\nabla^2\phi_N=4\pi G m\,\psi^*\psi.
\eea
We note that this is not simply the single particle Schr\"odinger equation, because the potential term $m\phi_N\psi$ is to be solved self-consistently with the Poisson equation, and so this system is in fact nonlinear. This accurately captures a system of bosons in the high occupancy condensed limit.
In fact the normalization of the wave function is the total number of particles $N=\int d^3x\,\psi^*\psi$.

{\em Orbiting Dwarf Galaxy}.---
Consider a dwarf galaxy in circular orbit around the center of a galactic halo with central location. Let the center of the dwarf be at a radius $a$ with orbital angular frequency $\omega=2\pi/T$; so Kepler's 3rd law gives $\omega^2=GM_{enc}/a^3$. Its center is taken to move clockwise in the $xy$-plane with location $x_c=a\,\cos(\omega t),\,\,y_c=-a\,\sin(\omega t),\,z_c=0$. 

Our goal is to compute the behavior of the bosons within the dwarf. It is therefore convenient to switch to a rotating coordinate system by defining
\bea
&&x'=x\,\cos(\omega t)-y\,\sin(\omega t),\\
&&y'=y\,\cos(\omega t)+x\,\sin(\omega t),\\
&&z'=z,\,\,\,\,\,\,
t'=t.
\eea
In these new coordinates, the $x'$ axis is always aligned from the halo center to the dwarf, i.e., $x_c'=a$ and $y_c'=z_c'=0$. We now insert these new coordinates into the Schr\"odinger-Poisson system, use the rotational symmetry of the Laplacian $\nabla'^2=\nabla^2$ and careful application of the chain rule to the time derivative, to obtain the slightly modified Schr\"odinger equation
\bea
&&i\,{\partial\psi\over\partial t'}=\omega\,\hat{L}_z\,\psi-{\nabla'^2\psi\over 2m}+m(\phidg+\phi_H)\psi,\\
&&\nabla'^2\phidg=4\pi Gm\psi^*\psi.
\eea
Here $\hat{L}_z=iy'{\partial\over\partial x'}-ix'{\partial\over\partial y'}$ is the angular momentum operator around the $z$-axis; this is a kind of Coriolis force term.
The Poisson equation is still structurally the same in these new coordinates, however, we have indicated that we break up the Newton potential into two pieces $\phi_N=\phidg+\phi_H$, where $\phidg$ is the potential from the dwarf, to be solved self-consistently, while $\phi_H$ is the potential from the halo, which we will treat as a fixed {\em external} potential. 

{\em Tidal Potential}.---
We treat the host halo as spherically symmetric for simplicity $\phi_H=\phi_H(r)$. The satellite is assumed to be much smaller in extent that the distance to the center $a$. This permits a Taylor expansion of the external potential $\phi_H$ in the vicinity of the halo as
\bea
&&\phi_H(\sqrt{(a+\tilde{x})^2+y'^2+z'^2})
=\phi_H(a)+\tilde{x}\,\phi_H'(a)\nonumber\\
&&\hspace{1.4cm}+{1\over2a}(a\,\tilde{x}^2\phi_H''(a)+(y'^2+z'^2)\phi_H'(a))+\ldots\,\,\,\,\,\,\,\,\,\,
\eea
where $\tilde{x}=x'-a$ is the distance from center of satellite along the rotated axis.
Truncating at quadratic order will suffice to obtain a tidal potential.

When inserted into the above Schr\"odinger equation, the linear term $\tilde{x}\,\phi_H'(a)$ seems to indicate we are not expanding around a local minimum of the potential. But in fact it simply represents the fact that we are expanding around an orbiting solution with angular momentum. We extract this out by removing a phase factor in the wave function as
\beq
\psi=e^{i\,m\,a^2\omega^2 t'-i\,m\,a\,\omega\, y'}\Psi,
\eeq
which absorbs the orbital $y'$ momentum $m\,a\,\omega$ of the dwarf's center. The resulting Schr\"odinger equation for $\Psi$ is found to be
(after dropping an irrelevant constant and for ease of notation we replace $\tilde{x}\to x,\,y'\to y,\,z'\to z,\,t'\to t$)
\beq
i\,{\partial\Psi\over\partial t}=\omega\,\hat{L}_z\,\psi-{\nabla^2\Psi\over 2m}+(m\phidg+ \Vt({\bf x}))\Psi,
\eeq
with $\nabla^2\phi_{DG}=4\pi Gm\Psi^*\Psi$. Here we have identified the tidal potential
\beq
\Vt({\bf x})=-{1\over2}m\,\omega^2(2\,\pf\, x^2-y^2-z^2).
\eeq
 The prefactor $\pf$ is a dimensionless property of the halo at the location of the dwarf
\beq
\pf\equiv-{a\,\phi_H''(a)\over2\,\phi_H'(a)}.
\label{HaloParameter}\eeq 
If the satellite is sufficiently far from the halo center that the enclosed mass $M_{enc}\approx M_{tot}$ is the total mass of the halo, then we know $\phi_H=-GM_{tot}/r$ and $\pf=1$. However, if we are still partially inside the halo, then $\pf<1$. In this work, we will often be interested in the case $\pf\approx1$, but we will also consider the more general case. 

We note that the corresponding single particle Hamiltonian $H=p^2/2m+\omega(xp_y-yp_x)+m\phidg+\Vt$ generates the classical equations of motion for a point particle.

{\em Mode Decomposition}.---
We see that this tidal potential is always highly aspherical; it gives rise to a potential instability along the $x$-axis between the halo and the dwarf, while it is confining in the orthogonal $yz$-plane. Furthermore the coefficient of the $x^2$ term is not parametrically larger than the $y$ or $z$ directions. So if we take a spherically symmetric ansatz for the wave function and integrate over angle, we will completely miss the tunneling. For instance, if $\pf=1$ and we integrate over solid angle, we have $\int d^2\Omega\,(2x^2-y^2-z^2)=0$, and the effects of the tidal term are lost.

To search for an exact solution, one should in principle, decompose the wave function into a sum of an infinite set of spherical harmonics
\bea
\Psi=e^{-i\,\mu\,t}\left[\Psi_0(r)+\sqrt{4\pi}\sum_{l>0,m}c_{lm}\Psi_{lm}(r) Y_{lm}(\theta,\varphi)\right]\!\!\!.\,\,\,\,\,\,\,\,\,
\label{FullModeAnsatz}\eea
These will be coupled to each other leading to an infinite chain of coupled mode functions; we return to the effects of this shortly. 
Here we have included a temporal phase factor in search of an eigenstate, with chemical potential $\mu$. A true stationary state has real $\mu$. But tunneling will be encoded in an imaginary part of $\mu$. This formally indicates exponential decay in the number $N$, however, this can be interpreted as the dwarf decaying and axions leaking out of the system.

To illustrate, let us describe a truncated ansatz for the solution by decomposing the wave function into a spherical piece (which will be dominant near the dwarf's center) and a quadrupolar $l=2$ piece to capture the quadratic aspherical potential (which will be comparable in the tail).
The theory carries an $x\to -x$ symmetry and so we anticipate the ground state to carry this too. So we set the $m=\pm 1$ terms to vanish; $c_{\pm1}=0$. Furthermore, we should extremize the integral over the tidal term. This means picking the quadrupole term to have maximal support in the $x$-direction and minimal support in the $yz$-plane. This occurs by taking 
$c_{2}=c_{-2}=\sqrt{3/8},\,c_0=-{1\over2}.$ 
(One can setup a straightforward extremization problem over these coefficients and check that indeed these values extremize the support in the $x$-direction versus the orthogonal plane.)
The normalization is $N=4\pi\int dr\,r^2(|\Psi_0|^2+|\Psi_2|^2)$. 

We then insert the above decomposition into an effective Hamiltonian which generates the above Sch\"odinger-Poisson system. We integrate over angle $\int d\Omega$ and then we need to extremize the Hamiltonian with respect to the radial mode functions $\Psi_0$ and $\Psi_2$. This leads to the following coupled system of radial ODEs
\bea
&&\mu\Psi_0=-{\nabla_r^2\Psi_0\over 2m}-{m\omega^2r^2\over2}(b_0\Psi_0+b_2\Psi_2)+m\bphidg\Psi_0,\,\,\,\,\,\,\,\,\,\,\label{Psi0Eq}\\
&&\mu\Psi_2=-{\nabla_r^2\Psi_2\over 2m}-{m\omega^2r^2\over2}(d_0\Psi_0+d_2\Psi_2)+m\bphidg\Psi_2\nonumber\\
&&\hspace{1.1cm}+{3\Psi_2\over m r^2}\label{Psi2},
\eea
where $\nabla_r^2\Psi = \Psi''+{2\over r}\Psi'$ designates the radial Laplacian and the final term in Eq.~(\ref{Psi2}) arises from the $l=2$ angular momentum of the $\Psi_2$ mode. The coefficients $b_0,\,b_2,\,d_0,\,d_2$ arise from carrying out the angular integrals of the tidal potential. (Their values in this truncated ansatz are $b_0={2\over3}(\pf-1),\,b_2=d_0=2(1+2\pf)/(3\sqrt{5}),\,d_2=2(11\pf-5)/21$.) Also, having integrated over angle, we can approximate the Newton potential for the dwarf as the solution to $\nabla_r^2\bphidg=4\pi G m(|\Psi_0|^2+|\Psi_2|^2)$. 

{\em Asymptotic Behavior and Weak External Potential.}---
Let us begin by examining the large $r$ regime of the above pair of equations. Here we know that the Newtonian potential will die off as $\sim 1/r$, etc, while the tidal term will be most important as it grows as $r^2$. In this regime, we can ignore all terms, except the tidal and Laplacian terms. The pair of coupled equations possess an oscillatory solution and a decaying solution. Since we are interested in tunneling, we can focus on the oscillatory one, which is found to be (also see Ref.~\cite{Hui:2016ltb} for the single mode case)
\beq
\Psi_0(r)=\alpha\,\Psi_2(r)\propto {1\over r^{3/2}}\exp\left(i{\sqrt{\beta}\,m\,\omega\over2}\,r^2\right)\,\,\,\,(\mbox{at large}\,\,\,r)
\label{larger}\eeq
with  
$\alpha=(3\sqrt{6}-\sqrt{5})/7$. 
This asymptotic behavior governs the tunneling. By inserting this relation between mode functions into Eq.~(\ref{FullModeAnsatz}) we can check the angular dependence of the solution for large $r$. 
The wave function in the truncated ansatz is self consistently large in the asymptotic regime along the $x$-axis and small in the $yz$-plane; in fact it is $\approx 58.5$ times smaller than along the $x$-axis. Ideally it would strictly vanish in both the $yz$-plane and at any ray that is not aligned with the $x$-axis, as this is the direction of tunneling.

More precisely then, taking into account self-rotation, and allowing for a full sum over spherical harmonics, one anticipates the value of the exponent in the tail to be
$\beta\approx 2\pf+1$; we shall use this going forwards.



For small $r$, we cannot use the above relation, and it appears we need to solve the full coupled system. However, there is a simplification that occurs when the tidal potential is small, as measured by $\omega$. For small $\omega$ and at small $r$, the theory is spherically symmetric and so we know $\Psi_2$ (and higher harmonics) will be negligible compared to $\Psi_0$. This suggests setting $\Psi_l=0$ for $l>0$ in the above equations. This suffices until we are at very large $r$ when we must track the corrections from the tidal term. But the latter we just solved above, finding a simple scaling in this regime ($\Psi_2=\Psi_0/\alpha$ in the truncated ansatz). So we can simply make this replacement in Eq.~(\ref{Psi0Eq}) knowing that it works precisely in the regime in which the relevant term matters, giving a single ODE. Taking this all into account gives
\beq
\mu\Psi_0=-{\nabla_r^2\Psi_0\over 2m}-{m\omega^2r^2\over2}\beta\,\Psi_0+m\bphidg\Psi_0.
\label{ODEsimp}\eeq
Furthermore, we can simplify the Poisson equation to $\nabla_r^2\bphidg=4\pi G m |\Psi_0|^2$, since in the weak tidal regime, we know that the integral of the spherical piece will dominate over the a-spherical piece in the bulk of the dwarf.
We note that for $\pf=1$ (point source halo regime) then 
$\beta\approx3$.
While deep within the halo $\pf$ is smaller and so too is $\beta$; we return to this later. 

{\em Numerical Results}.---
We have solved the above ODE (\ref{ODEsimp}) numerically for different values of the orbital frequency $\omega$ and halo parameter $\pf$. It is useful to define the dimensionless frequency
$\omegadim=\omega/\sqrt{G\,\rho_c}$,
where $\rho_c=m\,|\Psi_0(0)|^2$ is the central density.  In fact the only remaining tunable parameter is the combination $\beta\,\omegadim^2$ that appears in the tidal term. 

\begin{figure}[t]
   \centering
    \includegraphics[width=\columnwidth]{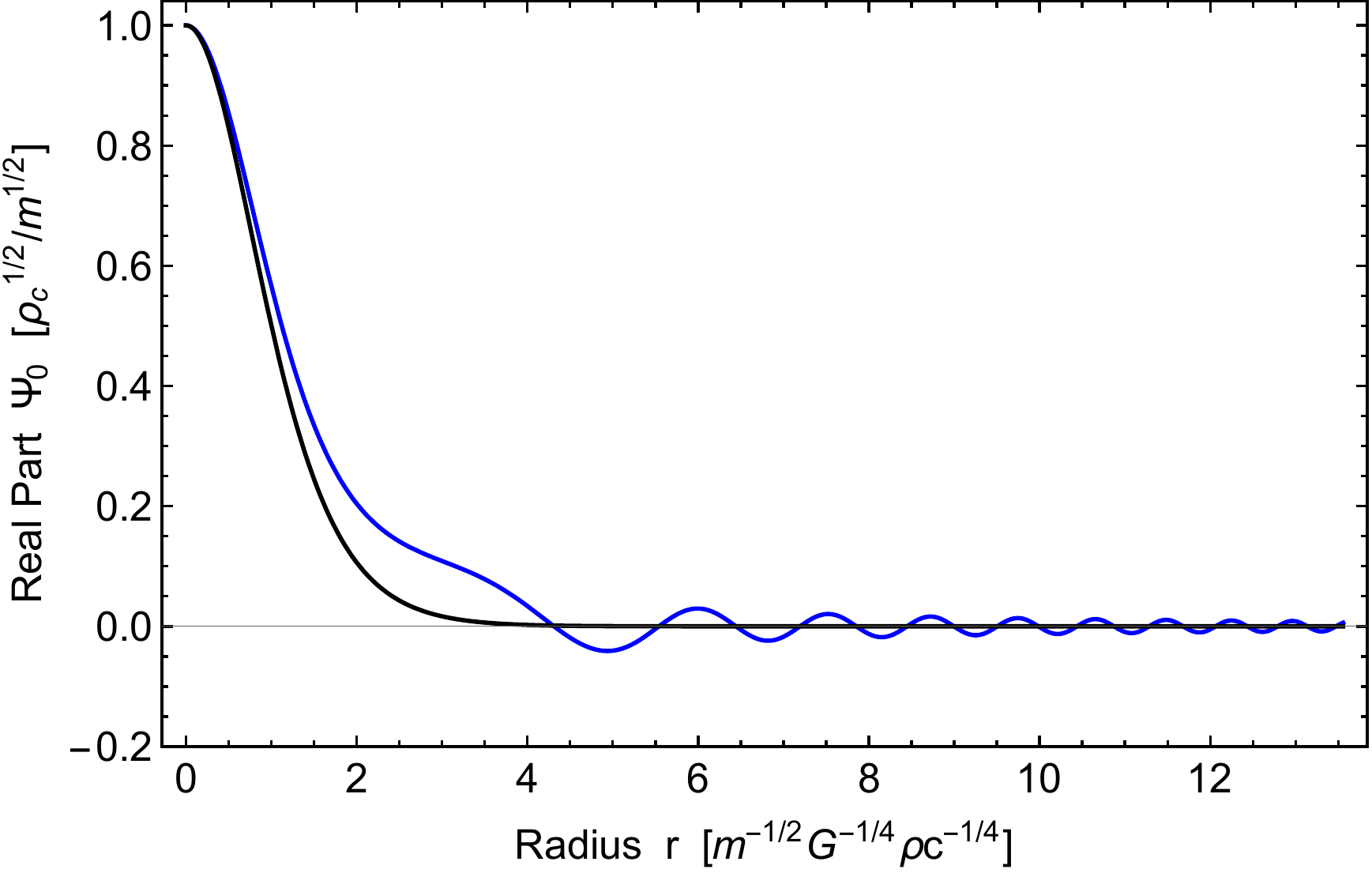}\\
    \includegraphics[width=\columnwidth]{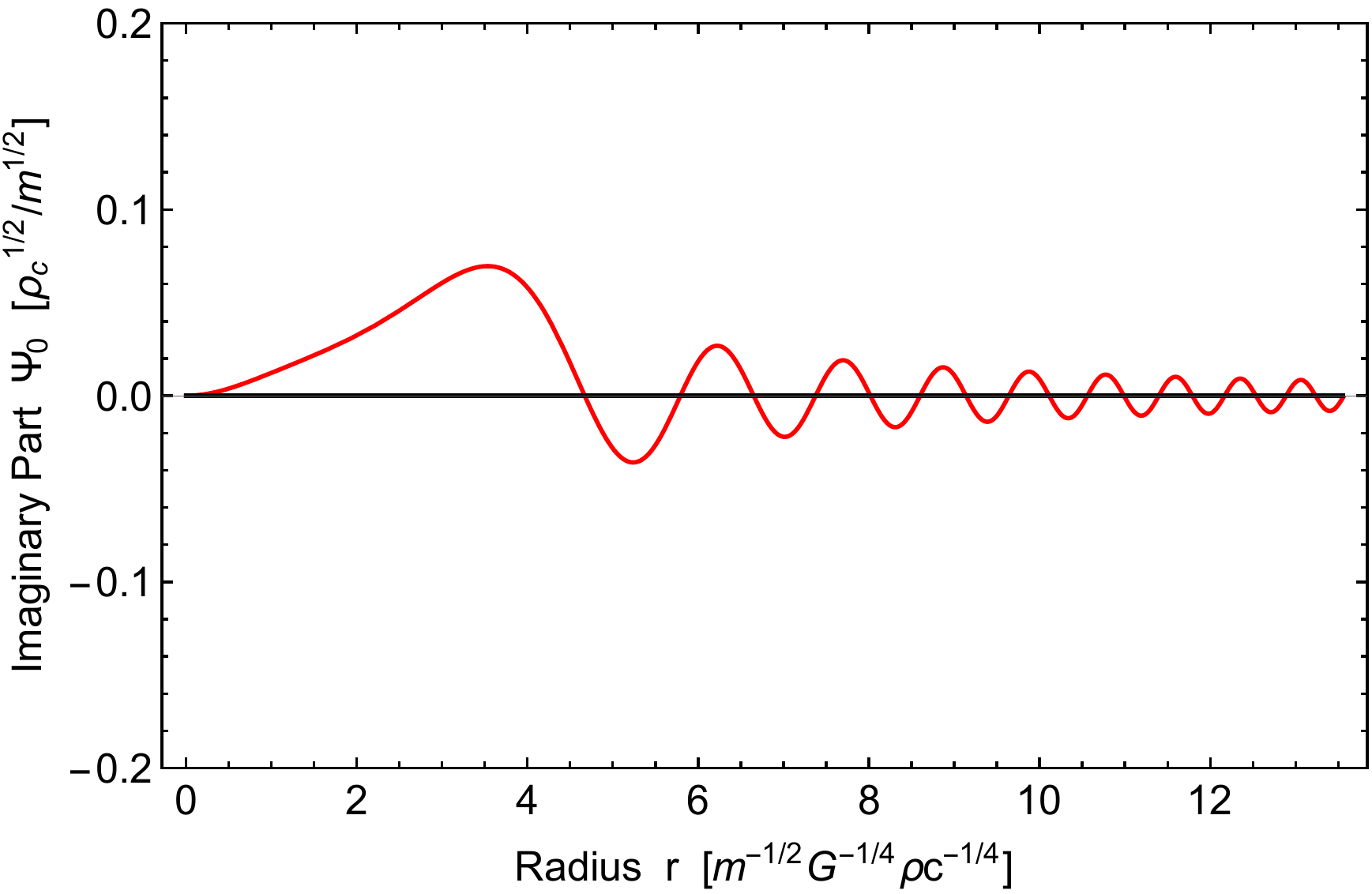}
   \caption{TOP: Real part of $\Psi_0$. BOTTOM: Imaginary part of $\Psi_0$. 
   Here $\omegadim=0.43,\,\pf=1$. 
   The solid blue (top) and solid red (lower) is from solving the radial equations numerically. 
   The black curve is the wave function without tidal force. 
   }
  \label{figWFradial}
\end{figure}

For $\pf=1$ and 
$\omegadim=0.43$, 
we display the result for the wave function versus radius in Fig.~\ref{figWFradial}. The upper plot gives the real part and the lower plot gives the imaginary part. We also indicate the unperturbed solution in black, which it matches well for small $r$. We have matched onto the large $r$ asymptotic behavior of Eq.~(\ref{larger}). 

This procedure requires numerically searching for the correct value of $\mu$, both real and imaginary parts $\mu=\mu_R+i\,\mu_I$. The imaginary part is related to decay. Since the wave function changes in time as $|\Psi|\propto e^{- |\mu_I]|t}$ (using $\mu_I<0$), then the number density changes as $n\propto e^{-2|\mu_I|t}$ and so in turn does the integrated mass $M=m\int d^3x\,n$ associated with the bound state. Naively, the corresponding instantaneous decay rate from tunneling is given by $\Gamma=|dM/dt|/M=2|\mu_{I}|$. However, as one tracks the adiabatic evolution of the soliton, as the density decreases, the radius increases as $R\propto e^{|\mu_I|t/2}$, so the net rate is altered to $\tilde{\Gamma}=|\mu_I|/2$. 

\begin{figure}[t]
   \centering
    \includegraphics[width=\columnwidth]{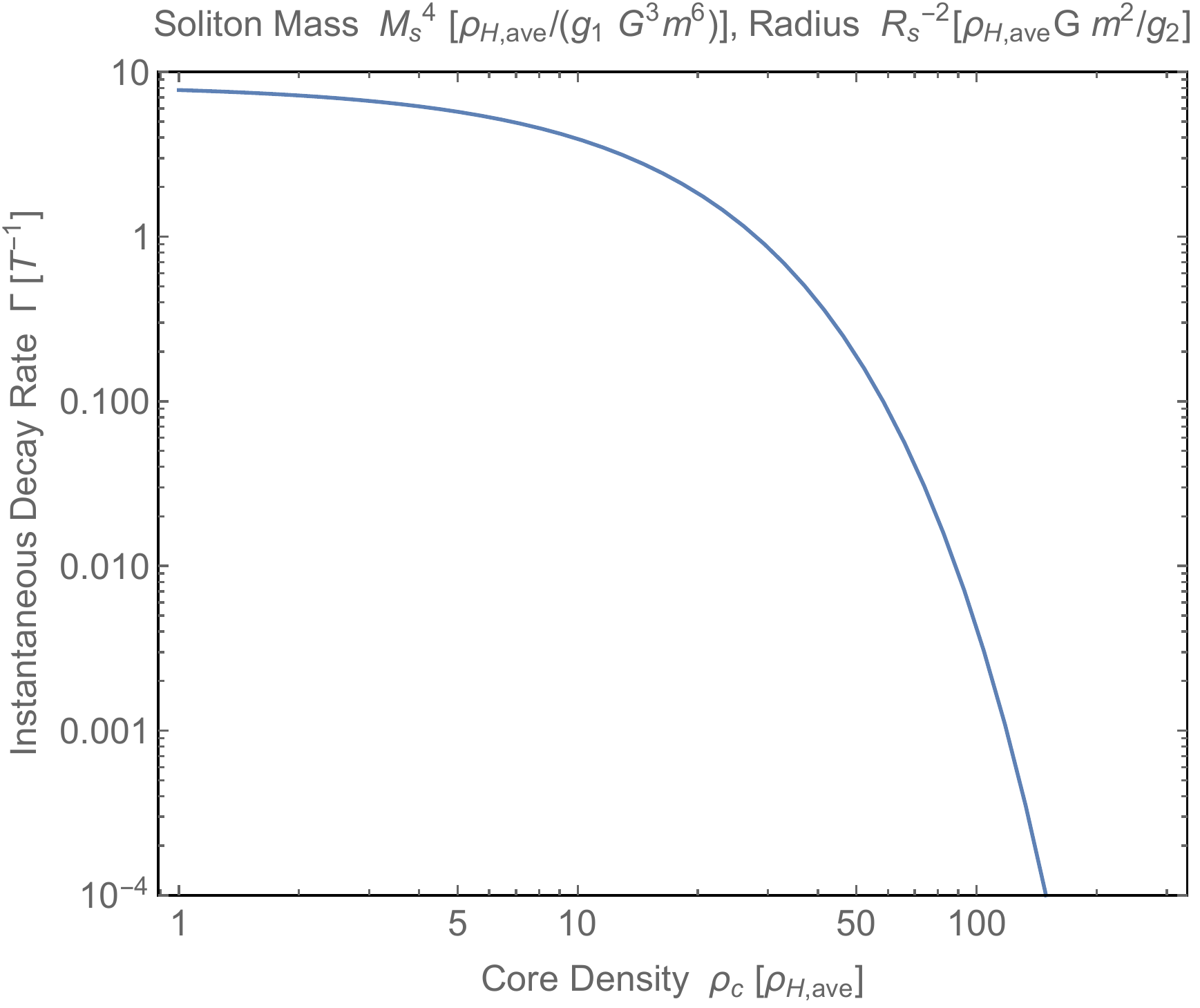}
   \caption{Instantaneous decay rate $\Gamma$ (in units of inverse orbital period $T^{-1}=\omega/(2\pi)$) as a function of dwarf central density $\rho_c$ (in units of average halo density $\rhoH=3\omega^2/(4\pi G)$), or equivalently as a function of soliton mass $M_s^4$ (in units of $\rhoH/(g_1\,G^3m^6)$) or soliton radius $R_s^{-2}$ (in units of $G m^2\rhoH/g_2$). Here $\gamma=1$.}
  \label{figLifetime}
\end{figure}

We can define a dimensionless density ratio $\rho_c/\rhoH=4\pi/(3\,\omegadim^2)$, where 
\beq
\rhoH={3\,\omega^2\over 4\pi G} = {3\,\pi\over G\,T^2}= {3\,M_{enc}\over 4\pi\,a^3}
\eeq
is defined as the density of the halo averaged to $a$ ($M_{enc}/V$). Then if we measure the decay rate in units of orbital frequency,  at fixed $\pf$ we obtain a unique curve. This is given in Fig.~\ref{figLifetime} for $\pf=1$.

Numerically, we find the approximate exponential fitting function for the decay rate 
\beq
\Gamma\approx \Gamma_0\exp\left(-\bc\,\rho_c/\rhoH\right)
\label{DecayRateFormula}\eeq
with 
\beq
\Gamma_0\approx 0.77 \sqrt{\beta}\,\omega,\,\,\,\,\, \bc\approx{0.23\over\beta}.
\eeq
(We obtain partial numerical agreement with Ref.~\cite{Du:2018qor}. However, unlike the formula in Ref.~\cite{Du:2018qor}, this has the physical properties that it is decreasing for arbitrarily large $\rho_c/\rhoH$ and is a function of the combination $\sqrt{\beta}\,\omega$ only).
Note that when expressed in these variables the axion mass $m$ has dropped out. This means that the observation of any long lived dwarf galaxy with a sufficiently small core density would falsify the proposal that its core is provided by condensed scalars of any mass $m$. 

In order to see an explicit dependence on axion mass, we need to switch variables. For sufficiently small $\omegadim$, the solution in the bulk is approximately that of the usual soliton (except of course in the large $r$ tunneling regime). The relation between core density $\rho_c$ and half mass of the soliton $M_s$ and radius containing half mass $R_s$ can be shown to be
\beq
\rho_c = g_1\,G^3m^6M_s^4 = {g_2\over Gm^2R_s^4}
\label{conversion}\eeq
where the dimensionless coefficients $g_{1,2}$ are found to be 
$g_1\approx 0.07$, 
$g_2\approx1.04$. By substituting this into the decay rate formula (\ref{DecayRateFormula}), we see a strong dependence on the axion mass and soliton mass $M_s$ or radius $R_s$. This is also indicated on the top axis of Fig.~\ref{figLifetime}.

{\em Application to Dwarfs and Bounds}.---
As an example, consider the ``Umi" spheroidal dwarf. This has a central density and orbital period of \cite{Read:2018fxs}
\beq
\rho_c\approx 0.15\,M_\odot/\mbox{pc}^{3},\,\,\,\,T\approx 1.6\,\mbox{Gyr}.
\eeq
This corresponds to the density ratio of $\rho_c/\rhoH\approx 183$. Then by  inserting this into the above tunneling rate formula with $\gamma=1$, we obtain
$\Gamma\sim 10^{-4}/t_{uni}$, 
where $t_{uni}\approx 13.8$\,Gyr is the current age of universe. This is a very small rate and therefore we would not expect this to be observable. 

On the other hand, for any dwarfs with a density ratio even a factor of a few smaller (either smaller core density or closer to host) would lead to appreciable tunneling on the lifetime of the universe. We can translate this into a bound
\beq
\rho_c/\rhoH\gtrsim 70.
\eeq
Any long lived dwarf with a density below this would falsify the soliton proposal. We can also convert this into a bound on the axion mass by using Eq.~(\ref{conversion}), which we find is
\bea
&&m\gtrsim 1.8\times 10^{-22}\,\mbox{eV}\left(4\times10^7\,M_\odot\over M_s\right)^{\!2/3}\!\left(3.9\,\mbox{Gyr}\over T\right)^{\!1/3}\,\,\,\,\label{mM}\\
&&m\lesssim 6.0\times 10^{-22}\,\mbox{eV}\left(0.71\,\mbox{kpc}\over R_s\right)^{\!2}\!\left(T\over 3.9\,\mbox{Gyr}\right),\label{mR}
\eea
where we have re-scaled the variables by reference values: soliton mass of $4\times10^7\,M_\odot$, period of 3.9\,Gyr, and radius of $R_s=0.71$\,kpc, which are the values for Fornax \cite{Read:2018fxs}. 
Note that the lower bound on $m$ arises from using the relationship between core density $\rho_c$ and the soliton mass $M_s$ in Eq.~(\ref{conversion}). While, the upper bound on $m$ arises from using the relationship between core density $\rho_c$ and the core radius $R_c$ in Eq.~(\ref{conversion}) and noting that this involves the axion mass {\em inversely}. This reflects the fact that soliton as core proposal is quite restrictive.
Our tunneling bound squeezes the axion mass into a narrow range. Any observed dwarf with a moderately smaller mass or moderately larger radius would falsify the proposal that axions in a condensed state form the cores. We also note that this upper bound is already in tension with a lower bound on the ultralight axion mass from Lyman-alpha forest \cite{Irsic:2017yje,Armengaud:2017nkf,Kobayashi:2017jcf,Rogers:2020ltq}. We note that Lyman-alpha bounds often assume the axion is all the DM or at least a substantial fraction of it, while our tunneling bounds do not need this requirement; only that axions provide an appreciable fraction of dwarf centers.


{\em Corrections Within Halo}.---
For satellites that are not entirely outside the halo, we can consider corrections from $\pf=1$. To model this we use an NFW profile for the halo as
\beq
\rhonfw(r)={\rho_0\over r/R_s(1+r/R_s)^2},
\eeq
where $\rho_0$ is central density of halo and $R_s$ is the ``scale radius" which is typically an order of magnitude smaller than the virial radius. The NFW density can be integrated to give the halo potential as
\beq
\phiH=-{4\pi G\rho_0 R_s^3\over r}\ln(1+r/R_s).
\eeq


For tunneling, we need the halo parameter defined earlier in Eq.~(\ref{HaloParameter}), which is found to be
\beq
\pf=1-{1\over 2(1+R_s/a)((1+R_s/a)\ln(1+a/R_s)-1)}.
\eeq
For $a\gg R_s$, we have $\pf\approx 1-1/(2\ln(a/R_s))$, while for $a\ll R_s$, we have $\pf\approx 2a/(3R_s)$.  
This has important implications for the tunneling parameter $\beta\,\omegadim^2$. 
At large radii $a\gg R_s$, the tunneling parameter is $\beta\,\omegadim^2\approx 12\pi(\rho_0/\rho_c)(R_s/a)^3\log(a/R_s)$, while at small radii $a\ll R_s$ the tunneling parameter is $\beta\,\omegadim^2\approx 2\pi(\rho_0/\rho_c)(R_s/a)$. So although the tunneling increases as we go to smaller radii, it only grows as $1/a$, rather than $1/a^3$. So sufficiently dense dwarfs could still be present at small radii.

{\em Discussion}.---
In this work we have focussed on the soliton state. However, as shown in Ref.~\cite{Deng:2018jjz}, the soliton's prediction for the relationship between core density and core radius over a range of galaxies does not fit the data well. Therefore, one should consider the possibility that a significant fraction (most) of the particles are {\em not} in the soliton state, but are instead in some higher energy excited state within the satellite, and that the core is provided by some other explanation. However, higher energy states are expected to tunnel even more quickly than our above estimates. This is expected to appreciably raise the lower bound on the axion mass in Eq.~(\ref{mM}), further restricting the ultralight axion as DM proposal.

For sufficiently small orbital radii, the system will be torn apart classically. However, for moderate to large radii the above pattern could be an interesting signature of ultralight DM. For a refined prediction, one should compute corrections from non-circular satellite orbits, which we leave for further work. 

It would be of interest to search for evidence of, or further constrain, satellites that obey the above mentioned tunneling rate as a function of radius.
Furthermore, it would be important to explore all possible dwarfs to see if the inequalities in Eqs.~(\ref{mM},\,\ref{mR}) are incompatible for some. If all dwarfs obey the inequalities, then it would be an intriguing success.
Any clues we can gather on the nature of DM is essential to progress in cosmology.



{\em Acknowledgments}.---
We thank Lam Hui and Matt Reece for helpful discussions.
M.~P.~H is supported in part by National Science Foundation grant PHY-2013953. 
The work of A.~L is supported in part by the Black Hole Initiative, which is funded by grants from the John Templeton Foundation and the Gordon and Betty Moore Foundation.

\appendix


\end{document}